\documentclass[aip,pop,amsmath,amssymb,reprint]{revtex4-2}

\usepackage{graphicx}
\usepackage{amsmath}
\usepackage{bm}
\usepackage{color}

\begin{document}

\title{Two-dimensional complex (dusty) plasma with active Janus particles}

\author{V. Nosenko}
\email{V.Nosenko@dlr.de}
\affiliation{Institut f\"{u}r Materialphysik im Weltraum, Deutsches Zentrum f\"{u}r Luft- und Raumfahrt (DLR), D-51147 Cologne, Germany}
\affiliation{Center for Astrophysics, Space Physics, and Engineering Research, Baylor University, Waco, Texas 76798-7310, USA}

\date{\today}
\begin{abstract}
A two-dimensional complex plasma containing active Janus particles was studied experimentally. A single layer of micron-size plastic microspheres was suspended in the plasma sheath of a radio-frequency discharge in argon at low pressure. The particle sample used was a mixture of regular particles and Janus particles, which were coated on one side with a thin layer of platinum. Unlike a suspension consisting of regular particles only, the suspension with inclusion of Janus particles did not form an ordered lattice in the experimental conditions used. Instead, the particles moved around with high kinetic energy in a disordered suspension. Unexpectedly, the mean kinetic energy of the particles declined as the illumination laser power was increased. This is explained by the competition of two driving forces, the photophoretic force and the oppositely directed ion drag force. The mean-squared displacement of the particles scaled as $t^{\alpha}$ with $\alpha=2$ at small times $t$ indicating ballistic motion and $\alpha=0.56\pm0.27$ at longer times due to the combined effect of the Janus particle propensity to move in circular trajectories and external confinement.
\end{abstract}


\maketitle

\section{Introduction}

A complex, or dusty plasma is a suspension of nanometer to micrometer size particles of solid matter in a gas-discharge plasma \cite{Ivlev_book}. The particles become charged due to the collection of electrons and ions from plasma and through their interaction and external confinement self-organize into liquid-like or solid-like structures. Complex plasmas are excellent model systems which allow studying of various plasma-specific and generic phenomena at the level of individual particles. Their advantages include the possibility of directly and relatively easily observing virtually undamped dynamics of the particles in real time. Due to the low neutral gas damping rate, the particle inertia becomes important, which distinguishes complex plasmas from such model systems as colloidal suspensions \cite{Ivlev_book}. Complex plasmas were successfully used to study transport phenomena \cite{Nunomura:2006,Nosenko:04PRL_visc,Gavrikov:2005,Hartmann:2011,Nosenko:08PRL_therm}, phase transitions \cite{Thomas:1996,Nosenko:2009,Melzer:2013}, as well as waves and instabilities \cite{Nunomura:2002,Piel:2002,Zhdanov:2003,Avinash:2003,Couedel:2010}.

Recently, the scope of complex plasmas as model systems was extended to include active matter systems. Active matter is a collection of active particles, each of which can extract energy from their environment and convert it into directed motion, thereby driving the whole system far from equilibrium \cite{Elgeti:2015,Bechinger:2016}. Active matter has some intriguing physical properties and potentially a number of applications in catalysis, chemical sensing, and health care. Recent research trends in the field of active matter include systems consisting of active particles with inertia \cite{Loewen:2020,Caprini:2021} and mixtures of active and regular (passive) particles \cite{Hauke:2020}. Complex plasmas are ideally suited for experiments in both of these subfields.

A particle in a complex plasma can become active, that is achieve self-propulsion, via several mechanisms. First, nonreciprocal interparticle interaction due to the plasma wake effect \cite{Melzer:1996,Lampe:2000,Ivlev:2015} can under certain conditions lead to the particle self-propulsion. Examples include channeling particles \cite{Du:2014} and spinning particle pairs (``torsions'') \cite{Nosenko:2015}. Second, a particle can be driven by a phoretic force, e.g., the photophoretic force from the illumination laser \cite{Du:2017,Wieben:2018}. Third, in extreme cases a particle can be propelled by the ``rocket force'' due to the ablation and removal of the particle material by a powerful laser irradiation \cite{Krasheninnikov:2010,Nosenko:2010}. It was recently shown that polymer microspheres coated on one side with a thin layer of platinum (the so-called Janus particles (JP) \cite{Walther:2013,Bechinger:2016}) become active when suspended in a radio-frequency (rf) argon plasma \cite{Nosenko:2020PRR_JP}. The emphasis was on the behaviour of single JPs, which were shown to be {\it circle swimmers} moving along characteristic looped trajectories. In Ref.~\cite{Arkar:2021}, single polymer microparticles partially coated with iron and suspended in an rf argon plasma were shown to move along complex jerky trajectories.

In this paper, we experimentally study a single-layer complex plasma composed of a mixture of regular melamine formaldehyde (MF) microspheres and active Janus particles similar to those used in Ref.~\cite{Nosenko:2020PRR_JP}. We find stark differences with a similar single layer consisting only of regular MF microspheres in the way of the its structure and dynamics.

\section{Experimental method}

The experiments described in this paper were carried out in a modified Gaseous Electronics Conference (GEC) radio-frequency (rf) reference cell \cite{Couedel:2022}. Plasma was produced by a capacitively coupled rf discharge in argon at $13.56$~MHz. The gas pressure was $p_{\rm Ar}=1.66$~Pa, the rf discharge power was $P_{\rm rf}=20$~W \cite{Hargis:1994}.

The particle sample used in our experiments was a mixture of regular melamine formaldehyde (MF) microspheres and active Janus particles. It was prepared using the method described in Ref.~\cite{Nosenko:2020PRR_JP}. MF microspheres \cite{microparticles} with a diameter of $9.19\pm0.09$~$\mu$m and mass $6.14\times10^{-13}$~kg were dispersed in isopropanol. A drop of the suspension was placed on a Si wafer and allowed to dry up. Unlike in Ref.~\cite{Nosenko:2020PRR_JP}, where the particles formed a monolayer on the wafer surface, here the amount of particles was larger and they formed a thicker layer. The wafer with particles was then placed in a sputter deposition machine and coated with a $\approx 10$~nm layer of platinum. Only the particles in the upper layer received the coating (on one side). Given this deposition technique, only a small fraction of all particles (a few percent) received metal coating resulting in their conversion into Janus particles. All particles were then separated from the wafer by a sharp blade.

The particles were injected into the plasma from a manual dispenser mounted in the upper flange. They were suspended in the plasma sheath above the lower rf electrode, where they formed a single layer. After injection, the particle suspension was cleaned using a standard procedure \cite{Du:2012}, where the discharge power was gradually reduced until larger particles and agglomerations of particles fell down to the rf electrode; the discharge power was then restored. The neutral-gas damping rate for the particles (which has the physical meaning of the collision frequency of the neutral gas atoms with particles) was calculated using the Epstein expression \cite{Epstein:1924} $\gamma=\delta N_gm_g\overline{v}_g(\rho_pr_p)^{-1}$, where $N_g$, $m_g$, and $\overline{v}_g$ are the number density, mass, and mean thermal speed of gas atoms and $\rho_p$, $r_p$ are the mass density and radius of the particles, respectively. With leading coefficient $\delta=1.39$ for the diffuse reflection of gas atoms from the particle, this gave $\gamma=2.12~{\rm s}^{-1}$.

The particles were illuminated by a horizontal laser sheet which had a Gaussian profile in the vertical direction with a standard deviation $\sigma\simeq75~\mu$m (corresponding to a full width at half maximum of $175~\mu$m) \cite{Couedel:2010}. The illumination laser had the wavelength $\lambda=660$~nm and variable output power of up to $100$~mW. The particles were imaged from above using the Photron FASTCAM mini WX100 camera equipped with the Nikon Micro-Nikkor $105$-mm lens fitted with a matched bandpass interference filter. This $4$-Megapixel, monochrome, $12$-bits per pixel camera has onboard memory of $16$~GB, which allowed recording of up to $2726$ frames. The camera frame rate was set to $125$ frames per second, resulting in the maximum recording duration of $21.8$~s. Experimental data were analysed in the following way. In each frame, the particle coordinates were calculated with subpixel resolution using a moment method \cite{SPIT}. Then individual particles were traced from frame to frame and their velocities were calculated from their displacements between frames.

\section{Results and discussion}

\begin{figure}[tb]
\centering
\includegraphics[width=1.0\columnwidth]{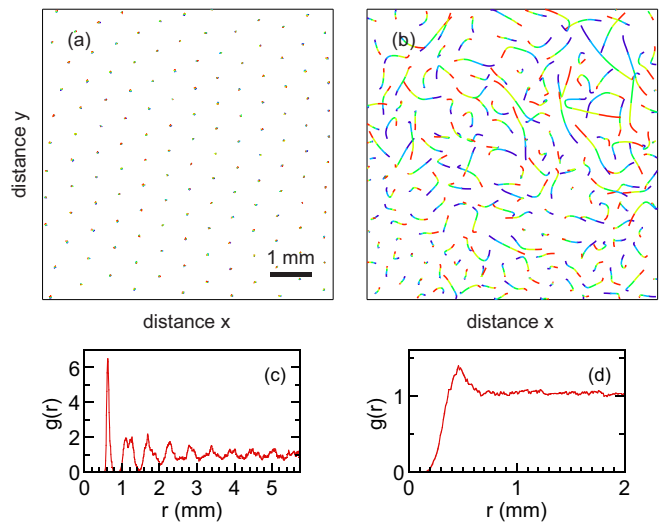}
\caption {\label {Fig_traj} Trajectories of (a) regular MF and (b) mixed Janus particles suspended as a single layer in rf plasma sheath, during $0.36$~s (time is color-coded from purple to red). Panels (c) and (d) show respective pair correlation functions $g(r)$. The argon pressure was $p_{\rm Ar}=1.66$~Pa, the rf discharge power was $P_{\rm rf}=20$~W, the illumination laser power was $P_{\rm laser}=14$~mW. The data were recorded after injecting the particles and cleaning the suspension.
}
\end{figure}

\begin{figure}[tb]
\centering
\includegraphics[width=0.86\columnwidth]{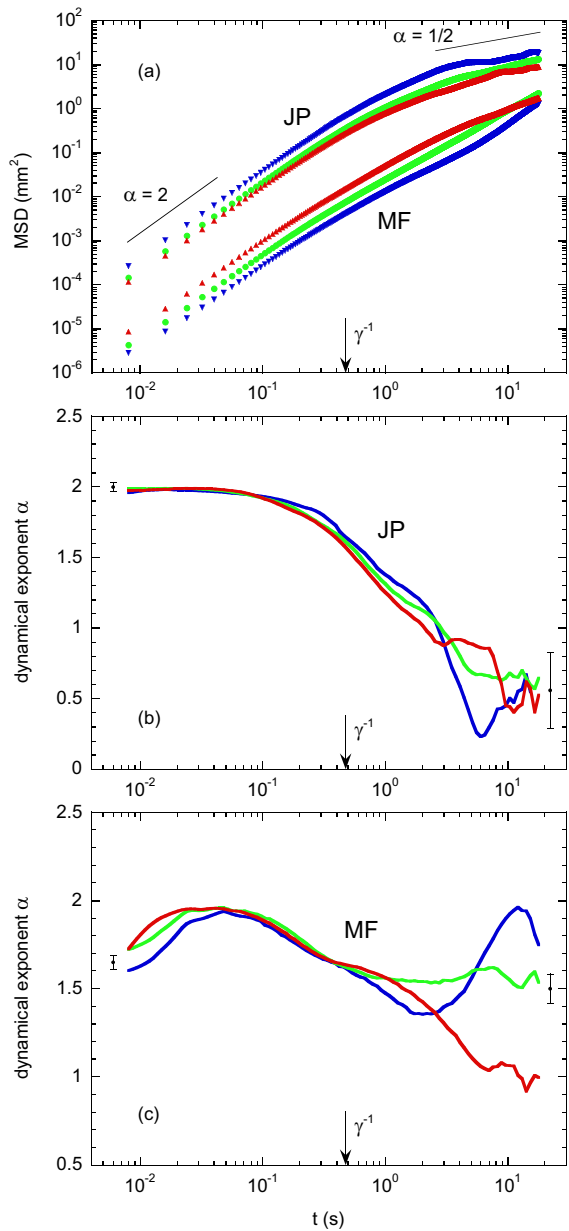}
\caption {\label {Fig_MSD} (a) Mean-squared displacement MSD(t) of the mixed Janus particles (upper curves) and regular MF particles (lower curves). (b),(c) Dynamical exponent $\alpha$ for the mixed Janus particles and regular MF particles, respectively. The left (right) error bars are for $t<2$~s ($t>2$~s). The illumination laser power was $P_{\rm laser}=14$~mW (blue down triangles and lines), $76$~mW (green circles and lines), and $99$~mW (red up triangles and lines).  The inertial delay time $\tau_m=\gamma^{-1}$ is shown by vertical arrows. The argon pressure was $p_{\rm Ar}=1.66$~Pa, the rf discharge power was $P_{\rm rf}=20$~W.
}
\end{figure}

As expected, the regular MF particles formed a two-dimensional triangular lattice (plasma crystal) in our experimental conditions (argon pressure $p_{\rm Ar}=1.66$~Pa, rf discharge power $P_{\rm rf}=20$~W), see Fig.~\ref{Fig_traj}(a). The lattice consisted of $\approx1900$ particles and was highly ordered, as evidenced by the pair correlation function for particles $g(r)$ with the high first and split second peaks, see Fig.~\ref{Fig_traj}(c). The lattice contained, however, a few energetic particles that locally disturbed it, which is similar to previous experiments, e.g. Refs.~\cite{Du:2017,Nosenko:2006}. These were most probably particles with slightly different sizes or irregular shapes, possibly damaged particles (called ``abnormal'' particles in Ref.~\cite{Du:2017}). They moved intermittently with high kinetic energy \cite{Nosenko:2006}, their trajectories were often irregular \cite{Du:2017}. They transferred a part of their kinetic energy to the neighboring particles via collisions. There were up to $10$ such ``active centers'' in the plasma crystal, which were distributed non-homogeneously (probably, due to the electric field inhomogeneity). Otherwise, the lattice was stable, in particular with respect to the mode-coupling instability (MCI) \cite{Couedel:2010}.

On the contrary, the suspension of mixed Janus particles did not crystallize. Instead, the particles energetically moved around colliding with each other, see Fig.~\ref{Fig_traj}(b). This is similar to the experiment of Ref.~\cite{Nosenko:2020PRR_JP}, where Janus particles moved around in characteristic curly trajectories and did not form an ordered lattice. The pair correlation function for mixed Janus particles $g(r)$, see Fig.~\ref{Fig_traj}(d), indicates a highly disordered (gas-like) state. These observations suggest that there must be some kind of energy input or external drive on the Janus particles. In Ref.~\cite{Nosenko:2020PRR_JP}, it was found that the individual Janus particles behave as {\it circle swimmers} when illuminated by a laser. The driving force on the Janus particles was identified as the photophoretic force caused by the illumination laser. In our experiment, the mixture of active Janus particles and passive MF particles appears rather homogeneous. The energy influx into the particle system due to the activity of Janus particles is effectively redistributed to the passive MF particles due to the interparticle interactions. Therefore, distinguishing between the two particle sorts is not straightforward.

To characterize the apparently random particle motion, we used their mean-squared displacement,
\begin{equation}\label{MSD}
{\rm MSD}(t)=\langle|{\bf r}_i(t)-{\bf r}_i(t_0)|^2\rangle,
\end{equation}
where ${\bf r}_i(t)$ is the position of the $i$-th particle at time $t$. The brackets denote the average over $10$ different times $t_0$ separated by $0.4$~s (i.e., $\simeq\gamma^{-1}$) and over all particles. The ${\rm MSD}(t)$ was measured for the whole particle suspension, which in the case of regular MF particles included ordered crystalline domains and also energetic particles. The ${\rm MSD}(t)$ of the mixed Janus particles as well as regular MF particles are shown in Fig.~\ref{Fig_MSD}(a). The mean-squared displacement of the mixed Janus particles scales as ${\rm MSD}(t)\propto t^{\alpha}$ with $\alpha=2$ at small times $t\ll\gamma^{-1}$ indicating ballistic motion. Here, the particle inertia is important due to the low gas damping rate. At later times, the dynamical exponent defined as \cite{Wang:2018,Hanes:2012}
\begin{equation}\label{alpha}
\alpha(t)=\frac{{\rm d\,ln}({\rm MSD}(t))}{{\rm d\,ln}(t)}
\end{equation}
declines, finally reaching the value of $\approx 0.56$, see Fig.~\ref{Fig_MSD}(b). ($\alpha$ was further smoothed using Stineman function, the error bars were calculated as the r.m.s. residuals of the respective fits.) We ascribe this to the combined effect of the Janus particle propensity to move in circular trajectories and external confinement. Note that no superballistic regime ($\alpha>2$) was observed.

The dynamical exponent $\alpha(t)$ for the regular MF particles is shown in Fig.~\ref{Fig_MSD}(c). It starts from a value below $2$, reaches $\simeq2$, and then declines. At later times, $\alpha(t)$ depends strongly on the illumination laser power $P_{\rm laser}$, varying from $\simeq2$ for the lowest $P_{\rm laser}$ to $\simeq1$ for the highest $P_{\rm laser}$. This behavior is due to the intermittent effect of the energetic particles. Since the energetic particles have much higher velocities than the particles in the crystalline areas, the MSD of the whole particle suspension is dominated by a few energetic particles and their immediate neighbors. Since the effect of the energetic particles is intermittent, the observed trends in MSD cannot be reliably extrapolated to longer times.

To compare our results with theory and computer simulations, we note that our experimental system is an ensemble of regular (passive) particles with an addition of small amount of active Langevin particles (self-propelled particles with inertia \cite{Loewen:2020,Caprini:2021,Sprenger:2021}) placed in a weak horizontal and strong vertical confinement (this situation is known as active doping \cite{Bechinger:2016}). The particles are thus confined to a plane with little out-of-plane motion (resulting in a quasi-2D system), but with 3D rotations. They interact with each other via a screened-Coulomb potential which is approximated reasonably well by the Yukawa potential \cite{Kompaneets_PhD}.

The simplest model of active particles with inertia is the active Ornstein-Uhlenbeck model \cite{Loewen:2022,Caprini:2021}. For a single active particle in a harmonic confinement it predicts that in general case the dynamical
exponent in MSD takes on the following values as time progresses \cite{Loewen:2022}: $\alpha(t)=2,4,3,1,0$. In our single-layer system of mixed Janus particles, $\alpha(t)$ declined from $2$ at small times to $\approx 0.56$ at the maximum recorded time of $17.8$~s, see Fig.~\ref{Fig_MSD}(b). The apparent lack of superballistic regime ($\alpha>2$) is probably explained by the small value of the dimensionless particle mass $\tilde{m}=\tau_m/\tau=(\gamma\tau)^{-1}$ in the present experimental conditions, due to the relatively large activity persistence time $\tau$, see Fig.~3(a) in Ref.~\cite{Loewen:2022}. At later times, $\alpha(t)$ showed signs of stabilization around the value of $\approx 0.56$, see Fig.~\ref{Fig_MSD}(b). Ref.~\cite{Loewen:2022} predicted an oscillatory regime at later times, where a particle would oscillate in the harmonic confinement; this regime is characterized by constant MSD and $\alpha=0$. In the present experiments, the oscillatory regime was probably suppressed by the collisions between particles, which interrupted the particle oscillations.

Crowded environment situation \cite{Bechinger:2016} provides the opposite limiting case for our experimental system. It is instructive to compare our results to the 2D random Lorentz gas model, where active particles move ballistically or diffuse through a random lattice of fixed repelling obstacles. This model provides an idealized description of dynamical systems consisting of two sorts of particles, one fast and the other slow. In Ref.~\cite{Morin:2017}, the dynamics of self-propelled colloidal rollers placed on a flat substrate with randomly distributed stationary repelling microposts was studied experimentally. The colloidal rollers were $4.8$-$\mu$m diameter polystyrene beads immersed in hexadecane and made motile by Quincke electrorotation, the microposts were produced by conventional UV lithography. Subdiffusive motion of particles was observed at later times and the dynamical exponent $\alpha$ declined with increasing density of the obstacles down to the values $\alpha\approx0$ indicating a localization transition. In Ref.~\cite{Voigtmann:2009}, a molecular-dynamics simulation of the equimolar binary mixture of purely repulsive soft-interacting spheres with the size ratio of $0.35$ and equal masses was performed. Smaller particles were observed to diffuse faster than the larger ones; they showed subdiffusive motion with long-time $\alpha=0.2$--$1$ and a localization transition at a higher density. In the present experiment, we observed subdiffusive motion of the mixed Janus particles with $\alpha=0.56\pm0.27$, but no localization transition. We ascribe this behaviour to the Janus particle propensity to move in circular trajectories \cite{Nosenko:2020PRR_JP} and to their external confinement.

The magnitude of ${\rm MSD}(t)$ of the mixed Janus particles is $1-2$ orders of magnitude larger than that of the regular MF particles, which indicates larger displacements and larger average speeds of the Janus particles, apparently due to their activity. For the regular MF particles, the magnitude of ${\rm MSD}(t)$ gets larger for higher illumination laser power. Unexpectedly, for the Janus particles the dependence is opposite: the ${\rm MSD}(t)$ magnitude gets smaller for higher laser power. We will address this finding in more detail below.

\begin{figure}[tb]
\centering
\includegraphics[width=0.9\columnwidth]{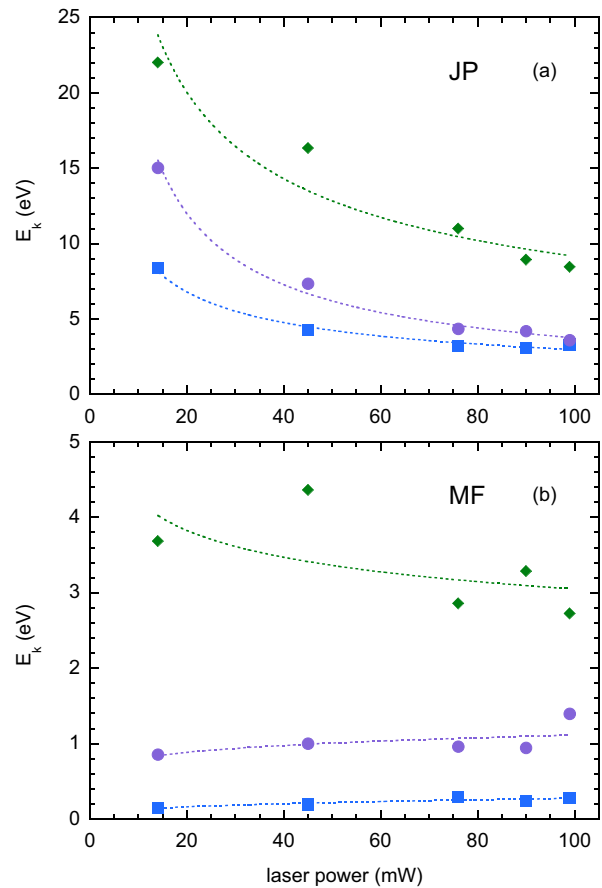}
\caption {\label {Fig_T} Mean kinetic energy $\langle E_k \rangle$ of (a) mixed Janus particles and (b) regular MF particles as a function of the illumination laser power measured at three different times: (blue squares) after injecting the particles and cleaning the suspension, (purple circles) after a waiting time of $40$~min, (green diamonds) after a waiting time of $180$~min. The power-law fits are shown to highlight the trends.
}
\end{figure}

To clarify the effect of the illumination laser power on the particle motion, we measured the mean kinetic energy of the particles $\langle E_k\rangle$ (averaged between all particles in the suspension) as a function of the illumination laser power at three different times: after injecting the particles and cleaning the suspension, after a waiting time of $40$~min, and after a waiting time of $180$~min. The results are shown in Fig.~\ref{Fig_T}. The mean kinetic energy $\langle E_k\rangle$ of mixed Janus particles indeed {\it decreases} when the laser power is increased at all measurement times, see Fig.~\ref{Fig_T}(a). On the other hand, $\langle E_k\rangle$ increases for longer waiting times. The mean kinetic energy of the regular MF particles increases with the laser power for the waiting times of $0$~min and $40$~min, see Fig.~\ref{Fig_T}(b). The increase of $\langle E_k \rangle$ is due to the increased total area of ``active centers''. In the crystalline and active areas themselves, the $\langle E_k \rangle$ does not in fact depend much on the laser power. For the longest waiting time of $180$~min, however, $\langle E_k \rangle$ decreases when the laser power is increased, see Fig.~\ref{Fig_T}(b).

The observed dependence of $\langle E_k\rangle$ on the illumination laser power can be explained in the following way. Two oppositely directed driving forces act on a Janus particle \cite{Nosenko:2020PRR_JP}: asymmetric ion drag force $F_i$ and the photophoretic force $F_{\rm ph}$. The ion drag force arises due to the momentum transfer from the ion flow in the vicinity of the particle and includes the collection and orbital parts \cite{Khrapak:2002,Nosenko:2007PoP}. We speculate that it is asymmetric for a Janus particle due to different electric properties of its Pt-coated and uncoated halves. Here, the component of $F_i$ {\it parallel} to the Janus particle axis of symmetry is considered, which comes on top of the main part of $F_i$, which is directed toward the rf electrode. The photophoretic force acts on a nonuniform object immersed in a neutral gas when their temperatures are not equal \cite{Mackowski:1989,Horvath:2014,Du:2017}. The Pt-coated side of a Janus particle is expected to have a higher temperature than the other side. Indeed, in the experiments of Ref.~\cite{Nosenko:2010}, MF particles with thin Pd coating absorbed the laser radiation more effectively than regular MF particles. Based on the observed dependence of $\langle E_k\rangle$ on $P_{\rm laser}$, we conjecture that $F_i>F_{\rm ph}$ for the Janus particles in the experimental conditions of the present work. Therefore, when the laser power is increased and $F_{\rm ph}$ becomes larger, the net force declines. For the regular MF particles, the driving force reduces to the photophoretic force only \cite{footnote,Soong:2010}, leading to the weakly rising dependence of $\langle E_k\rangle$ on the illumination laser power for the waiting times of $0$~min and $40$~min, see Fig.~\ref{Fig_T}(b). For the longest waiting time of $180$~min, however, the dependence becomes falling similarly to the mixed Janus particles. Whether the proposed model or some other particle propulsion mechanism (e.g., preferential plasma sputtering of one of the particle sides) is at work can be verified in future experiments, for example by looking at the scaling of the particle self-propulsion force with the gas pressure, discharge power, and the particle size.

The temporal variation trend of $\langle E_k\rangle$ may be due to the {\it in-situ} plasma deposition of a non-uniform patchy metal film on the surface of suspended particles similar to that observed in Ref.~\cite{Kononov:2021}. The acquired coating would in fact produce an imperfect Janus particle \cite{Kononov:2021}, leading to the falling dependence of $\langle E_k\rangle$ on the illumination laser power for the MF particles for the longest waiting time of $180$~min. For both regular MF and mixed Janus particles, $\langle E_k\rangle$ increased for longer waiting times, presumably because all suspended particles received more {\it in-situ} metal coating with time. Another reason of the gradually rising $\langle E_k\rangle$ may be continuing damaging of the particle surface due to plasma sputtering.

To summarize, a system consisting of micron-size melamine formaldehyde microspheres, some of which were coated on one side with a thin layer of platinum (Janus particles) and suspended as a single layer in an rf argon plasma was studied experimentally. Due to self-propulsion of the Janus particles the system became active and did not form an ordered lattice, unlike a similar system without inclusion of Janus particles in the same experimental conditions. The mean kinetic energy of the particles depended on the illumination laser power and the time the particles spent suspended in plasma. The dynamical exponent $\alpha$ of the particle mean-squared displacement declined from $\alpha=2$ at small times indicating ballistic motion to $\alpha=0.56\pm0.27$ at longer times due to the combined effect of the Janus particle propensity to move in circular trajectories and external confinement. No superballistic regime with $\alpha>2$ was observed. The experimental findings can be explained by an interplay between two oppositely directed driving forces acting on a Janus particle, asymmetric ion drag force and the photophoretic force.

\section{Acknowledgments}

Thomas Voigtmann is acknowledged for carefully reading the manuscript and helpful discussions.

\section{Author declarations}

The author has no conflicts of interest to disclose.

\section{Data availability}

The data that support the findings of this study are available from the corresponding author upon reasonable request.

\end{document}